\documentclass[preprint,showpacs,showkeys,amsmath,amssymb,aps,doublespace]{revtex4}
\usepackage[dvips]{graphicx}
\usepackage{setspace}
\usepackage{amsmath}
\usepackage{amsfonts}
\usepackage{amssymb,color}

\usepackage{graphicx,color}
\usepackage[pdftex]{hyperref}
\usepackage{ifpdf}
\usepackage[latin1]{inputenc}

\begin{document}

\title{Contact process with sublattice symmetry breaking}

\author{Marcelo Martins de Oliveira$^1$\footnote{email: mmdeoliveira@ufsj.edu.br} and Ronald Dickman$^2$\footnote{email: dickman@fisica.ufmg.br}
}
\address{
$^1$Departamento de F\'{\i}sica e Matem\'atica,
CAP, Universidade Federal de S\~ao Jo\~ao del Rei, \\
36420-000 Ouro Branco, Minas Gerais - Brazil\\
$^2$Departamento de F\'{\i}sica and
National Institute of Science and Technology for Complex Systems,\\
ICEx, Universidade Federal de Minas Gerais, \\
C. P. 702, 30123-970 Belo Horizonte, Minas Gerais - Brazil
}

\date{\today}

\begin{abstract}

We study a contact process with creation at first- and second-neighbor
sites and inhibition at first neighbors, in the form of an
annihilation rate that increases with the number
of occupied first neighbors. Mean-field theory predicts
three phases: inactive (absorbing), active symmetric, and active asymmetric,
the latter exhibiting distinct sublattice densities
on a bipartite lattice. These phases are separated by continuous
transitions; the phase diagram
is reentrant. Monte Carlo simulations in two dimensions verify
these predictions qualitatively, except for a first-neighbor creation rate of zero.
(In the latter case one of the phase transitions is {\it discontinuous}.)
Our numerical results confirm that the symmetric-asymmetric
transition belongs to the Ising universality class, and that the
active-absorbing transition belongs to the directed percolation class, as expected
from symmetry considerations.

\end{abstract}


\maketitle

\section{Introduction}

An absorbing-state phase transition is a far from equilibrium transition between
an active and an inactive phase.
When a control parameter such as a creation or annihilation rate is varied,
the system undergoes a phase transition
from a fluctuating state to an absorbing one.
Such transitions are currently
of great interest, arising in a wide
variety of problems, such as population dynamics,
heterogeneous catalysis, interface growth, and epidemic
spreading \cite{marro,henkel,odor07,hinrichsen,odor04}. Interest on this class of phase
transitions has been stimulated by recent experimental realizations \cite{take07,pine}.

As equilibrium statistical mechanics, absorbing-state phase transitions
can be classified into universality classes. However, a complete classification of their
critical behavior is still lacking; much numerical and theoretical work
is focused on identifying the factors that determine the universality classes of these transitions.

The absorbing-state universality class associated with directed percolation (DP) has proven
to be particularly robust.  DP-like behavior appears to be generic for absorbing-state
transitions in models with short-range interactions and lacking a conserved density
or symmetry beyond translational invariance \cite{gras-jans}.
By contrast, models possessing two
absorbing states linked by particle-hole symmetry belong to the voter model universality class \cite{voter}.
There are also models which, while free of absorbing states, cannot achieve thermal equilibrium
because their transition rates violate the detailed balance principle.  An example is the
majority vote model.  In its ordered phase a
$Z_2$ symmetry is spontaneously broken, leading to Ising-like behavior in two or more dimensions.

In this work we examine whether a spatially structured population model
that suffers a phase transition to a {\em single} absorbing state
can also exhibit a broken-symmetry phase. Our candidate for such
a model is a modified contact process (CP) with suppression of activity
at the nearest neighbors of active sites. We observe
unequal sublattice populations for suitable choices of the control
parameters.

The balance of this paper is organized as follows. In the next
section we define the model and analyze its mean-field theory.
In Sec. III we present our simulation results; Sec. IV is
devoted to discussion and conclusions.

\section{Model and Mean-Field Theory}

The CP \cite{harris-CP} is a stochastic interacting particle
system defined on a lattice, with each site $i$ either occupied by a particle
[$\sigma_i (t)= 1$], or vacant [$\sigma_i (t)= 0$]. Transitions from
$\sigma_i = 1$ to $\sigma_i = 0$ occur at a rate of unity,
independent of the neighboring sites. The reverse transition, at a vacant site, is only
possible if at least one of its neighbors is occupied: the
transition from $\sigma_i = 0$ to $\sigma_i = 1$ occurs at rate
$\lambda r$, where $r$ is the fraction of nearest neighbors of site
$i$ that are occupied. In the absence of spontaneous creation of particles,
the state $\sigma_i = 0$ for all $i$ is
absorbing. It can be shown \cite{harris-CP} that for a certain critical value
$\lambda_c$ the system undergoes a phase transition
between the active and the absorbing state. Although no exact results are
available, the CP has been studied intensively via series expansion and
Monte Carlo simulation, and its critical properties are known to high precision \cite{marro,henkel,odor07,hinrichsen,odor04}.

In order to generate distinct sublattice occupations, we modify the
basic contact process in two respects:

i) In addition to creation at first neighbors, at rate $\lambda_1$, we allow
creation at {\it second neighbors}, at rate $\lambda_2$.  For bi-partite lattices such as the
square or simple cubic lattice, $\lambda_1$ is the rate of creation in the
opposite sublattice, while $\lambda_2$ is the rate in the
same sublattice as the replicating particle.  Unequal sublattice occupancies
are thus favored for $\lambda_2 > \lambda_1$.

ii) We include a nearest-neighbor inhibition effect in the annihilation
process.  In addition to the intrinsic annihilation rate of unity, there
is a contribution of $\mu n_1^2$, where $n_1$ is the number of occupied
first neighbors.  Thus if the occupation fraction $\rho_A$ of sublattice A
is much larger than that of sublattice B, any particles created in sublattice B
will die out quickly, stabilizing the unequal sublattice occupancies.

The one-site mean-field theory (MFT) for this model on a lattice of coordination number $q$
is given by the coupled equations,

\begin{equation}
\frac{d \rho_A}{dt} = - (1 + \mu q^2 \rho_B^2) \rho_A + (\lambda_1 \rho_B + \lambda_2 \rho_A)
(1-\rho_A)
\end{equation}
and
\begin{equation}
\frac{d \rho_B}{dt} = - (1 + \mu q^2 \rho_A^2) \rho_B + (\lambda_1 \rho_A + \lambda_2 \rho_B)
(1-\rho_B)
\end{equation}

\noindent which are seen to be symmetric under $\rho_A \leftrightharpoons \rho_B$.  Defining
$\rho = \rho_A + \rho_B$ and $\phi = \rho_A - \rho_B$, these equations may be recast as,

\begin{equation}
\frac{d \rho}{dt} = (\Lambda - 1) \rho - \frac{\Lambda}{2} \rho^2 - \frac{\Delta}{2} \phi^2
-\frac{1}{4} \mu q^2 (\rho^2 - \phi^2) \rho
\label{drhodt}
\end{equation}
and
\begin{equation}
\frac{d \phi}{dt} = \left[ \Delta - 1  - \lambda_2 \rho -\frac{1}{4} \mu q^2 (\rho^2 - \phi^2)
\right] \phi
\label{dphidt}
\end{equation}
where $\Lambda \equiv \lambda_1 + \lambda_2$ and $\Delta \equiv \lambda_2 - \lambda_1$.

From Eq. (\ref{drhodt}) it is evident that at this level of approximation, the extinction-survival
transition occurs at $\Lambda = 1$, as one would expect.  This transition
leads to the symmetric active phase, characterized by $\rho > 0$ and $\phi = 0$.  This phase is
linearly stable for small $\Delta$.
As $\lambda_2$ is increased, with $\lambda_1$ constant, the symmetric active
phase can become unstable, leading to a phase with $|\phi| > 0$.
On further increasing $\lambda_2$, however, $\rho $ can increase to the point that
the symmetric active phase is again stable. (Perfect sublattice ordering corresponds to $\rho=\phi$,
which is only possible for $\rho \leq 1$;
total densities larger than unity tend to suppress sublattice ordering.) Thus the asymmetric
active phase should exist for some intermediate range of $\lambda_2$ values.

In the symmetric active phase, the stationary activity density is,
\begin{equation}
\rho=\frac{1}{2\kappa}\left[\sqrt{(\Lambda/2)^2+4\kappa(\Lambda-1)}-(\Lambda/2) \right]
\label{eqstat1}
\end{equation}
with $\kappa \equiv \mu q^2/4$. From Eq. (\ref{dphidt}), this solution is stable
with respect to one with $\phi\neq 0$, when
\begin{equation}
a_\phi \equiv \Delta-1-\lambda_2\rho+\kappa \rho^2 < 0.
\end{equation}

\noindent The transition to the asymmetric phase occurs when $a_\phi=0$, which implies
\begin {equation}
(2\lambda_2+\Lambda)^2(\Lambda-1)=2(\lambda_2-1)\left[8\kappa(\lambda_2-1)+(2\lambda_2+\Lambda) \Lambda \right]
\label{eqstat2}
\end{equation}

\begin{figure}[!hbt]
\includegraphics[clip,angle=0,width=0.8\hsize]{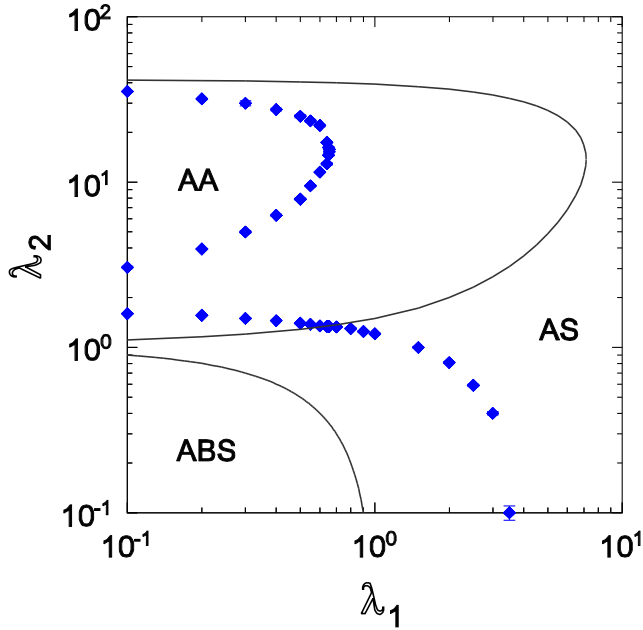}
\caption{\footnotesize{(Color online) Phase diagram in the $\lambda_1 - \lambda_2$ plane
for $\mu = 2$, showing absorbing (ABS), active-symmetric (AS) and active asymmetric (AA) phases.
Lines: MFT; symbols: simulation.}}
\label{cpslm2}
\end{figure}

Eq.~(\ref{dphidt}) may be written as $d\phi/dt=a_\phi\phi-\kappa\phi^3$, showing the expected symmetry
under $\phi\to -\phi$,
which in turn suggests that the transition between symmetric- and asymmetric-active phases is Ising-like.
The two transitions (active/absorbing and symmetric/asymmetric) coincide at the point
$\lambda_1=0$, $\lambda_2=1$.
Figure~\ref{cpslm2} shows the phase diagram for $\mu=2$, obtained using Eq. (\ref{eqstat2}).
Of particular note are the observations that (1)
the asymmetric-active phase is observed only for intermediate values of $\lambda_2$, i.e.,
the phase diagram is reentrant, and (2) the asymmetric-active phase does not exist for
$\lambda_1$ above a certain value, $\lambda_1^* (\mu)$.
For $\mu=2$, for example, $\lambda_1^* = 7.1443$.
Also shown in Fig.~\ref{cpslm2} are simulation results (details of our simulation method are discussed in the
following section).  MFT and simulation are in qualitative agreement, but as is usually the
case, MFT overestimates the regions in parameter space corresponding to the
active and ordered phases.  In particular, MFT overestimates $\lambda_1^*$ by about an order
magnitude.

The line $\lambda_1=0$ is special, since the subspaces with $\rho_A > 0$ and
$\rho_B = 0$ (and vice-versa), are also absorbing.  MFT yields the following
phases: absorbing for $\lambda_2 \leq 1$; asymmetric-active-I for $1 < \lambda_2 < \lambda^I(\mu)$;
asymmetric-active-II for $\lambda^{I}(\mu) < \lambda_2 < \lambda^{II}(\mu)$; and symmetric-active
for $\lambda_2 > \lambda^{II}$.  The difference between phases asymmetric-active-I and II is that
in the former, $\rho = \phi$ (i.e., only one of the sublattices is populated) while in the
latter $\rho > \phi > 0$.  The transitions between these phases are continuous.  (Naturally,
the asymmetric-active-II and symmetric-active phases can only be reached from initial conditions
with {\it both} sublattices populated.)  For $\mu=2$, for example, we find
$\lambda^{I} = 30.97$ and $\lambda^{II} = 41.65$.

We also studied a version in which the suppression (enhanced annihilation rate) is a {\it linear}
function of the number of occupied first neighbors, i.e., the mean-field equations

\begin{equation}
\frac{d \rho_A}{dt} = - (1 + \mu q \rho_B) \rho_A + (\lambda_1 \rho_B + \lambda_2 \rho_A)
(1-\rho_A),
\end{equation}

\noindent and similarly for $\rho_A \leftrightharpoons \rho_B$.  The phase diagram
is qualitatively similar to that found above, but the region occupied by the asymmetric-active
phase is much reduced.  For $\mu=2$, for example, sublattice ordering is only found
for $\lambda_1 \leq 0.6805$.
Since mean-field theory offers at best a qualitative description of the model, we turn, in the
following section, to simulation.

\section{Simulations}

We performed extensive Monte Carlo simulations of the model on square
lattices of linear size $L=20, 40,..., 320$ sites, with periodic boundaries.
The simulation algorithm is as follows. First, a
site is selected at random. If the site is occupied, it creates a particle at one of its
first-neighbors with a probability $p_1=\lambda_1/(1+\lambda_1+\lambda_2+\mu n_1^2)$, or
at one of its second-neighbors with a probability $p_2=\lambda_2/(1+\lambda_1+\lambda_2+\mu n_1^2)$.
With a complementary probability $1-(p_1+p_2)$ the site is vacated.
(In order to improve efficiency the sites are chosen from a list which contains the currently
$N_{occ}$ occupied sites; we increment the time by $\Delta t= 1/N_{occ}$ after each event).
For simulations in the subcritical and critical absorbing regime,
we employ the quasi-stationary (QS) simulation method detailed in \cite{qssim},
to further improve efficiency.

\begin{figure}[h]
\label{exponent}
\includegraphics[clip,angle=0,width=0.5\hsize]{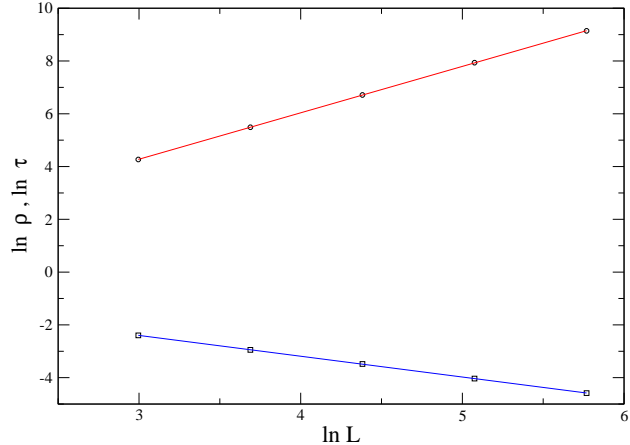}
\caption{{\footnotesize (Color online) Scaled critical QS density of active sites ln $\rho$ (bottom)
and scaled lifetime of the QS state ln $\tau$ (top), {\it
versus} ln$L$. Parameters: $\mu=2.0$, $\lambda_1=0.2$ and $\lambda_2=1.5620$}}
\end{figure}

\begin{figure}[h]
\includegraphics[clip,angle=0,width=0.5\hsize]{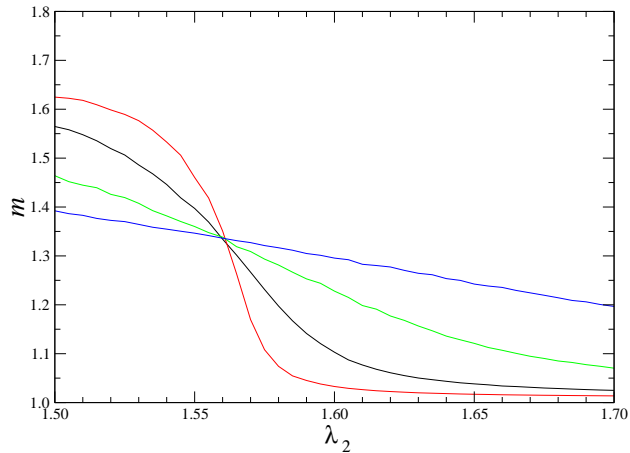}
\caption{{\footnotesize (Color online) Quasistationary moment ratio $m$ versus $\lambda_2$, for
$\mu=2.0$ and $\lambda_1=0.2$.
System sizes: $L=20,40,80,160$ (in order of increasing steepness).}}
\label{mabs}
\end{figure}

In a series of studies using $\mu=2.0$, we verify that for low values of $\lambda_1$ and $\lambda_2$ the system
becomes trapped in the absorbing (inactive) state.  When we increase $\lambda_1$
and/or $\lambda_2$ the system undergoes a phase transition from the inactive to the active
symmetric (AS) phase.
For example, for $\lambda_1=0.2$, the absorbing transition occurs at $\lambda_2=1.5620(5)$.
 (Note that at this point $\lambda_1 + \lambda_2 $ is somewhat greater than the
critical value for the basic process, $\lambda_c = 1.6488(1)$, on the square lattice \cite{agmrd}.)
At the critical point, the quasistaionary order parameter is expect to decay as a power
law, $\rho \sim L^{-\beta/\nu\perp}$. In Fig.~\ref{exponent}, we show that our data follow a power law,
with $\beta/\nu_\perp=0.79(1)$, in good agreement with the DP value
$\beta/\nu_\perp=0.797(3)$ \cite{dickman99}. The lifetime of the QS state
$\tau \sim L^{\nu_{||}/\nu_\perp}$ at criticality, where we find $\nu_{||}/\nu_\perp=1.75(2)$,
also shown in Fig.~2. This value again is in very good agreement with the best known DP
value $\nu_{||}/\nu_\perp=1.7674(6)$.  The moment ratio $m=\langle\rho^2\rangle/\langle\rho\rangle^2$
(see Fig.~\ref{mabs}) is analogous  to Binder's fourth cumulant \cite{bind81} at an equilibrium critical point,
in the sense that $m$ assumes a universal value $m_c$ at the critical point. We find $=1.324(5)$,
in comparison to the known DP value $m=1.3264(5)$ \cite{dic-jaf}.
These results permit us to state that, as expected, the absorbing transition belongs to the
universality class of directed percolation.

\begin{figure}[h]
\includegraphics[clip,angle=0,width=0.5\hsize,height=0.5\hsize]{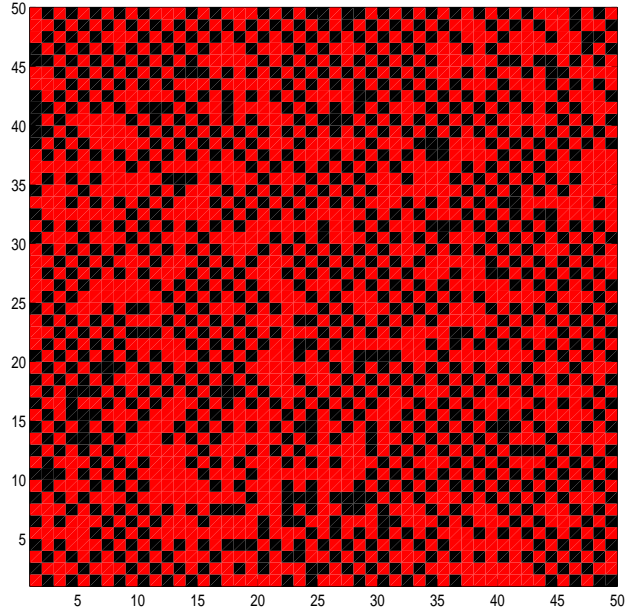}
\caption{\footnotesize{(Color online) A typical configuration in the active asymmetric phase;
particles are represented by black sites. System size: $L=50$. Parameters: $\mu=2.0$,
$\lambda_1=0.2$, and $\lambda_2=6.0$.}}
\label{cpsl6}
\end{figure}

\begin{figure}[h]
\label{dist}
\includegraphics[clip,angle=0,width=0.6\hsize]{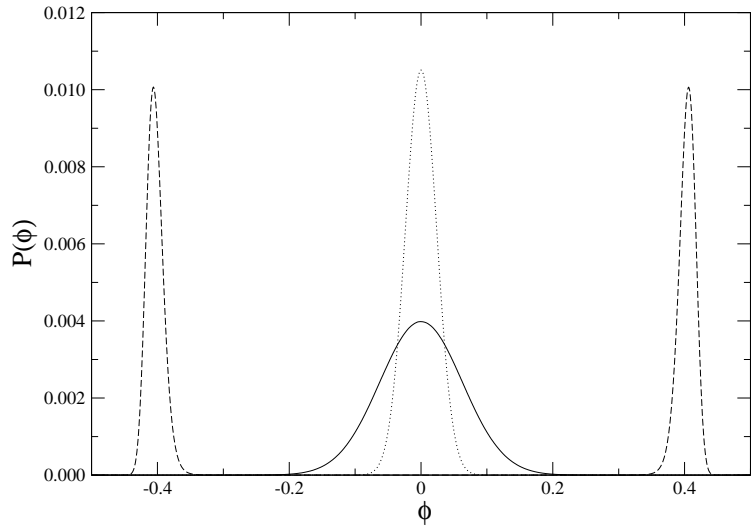}
\caption{{\footnotesize Probability distribution of the order parameter $\phi$.
System size: $L=80$. Parameters: $\mu=2.0$, $\lambda_1=0.2$.
$\lambda_2=2.0$(dotted curve), $10.0$ (dashed) and $40.0$ (solid).}}
\end{figure}

In the active phase, we find that as we increase $\lambda_2$ at fixed $\lambda_1$,
there is a transition to a state with unequal particle densities in the two
sublattices ($\phi \neq 0$), that is, to
the active asymmetric (AA) phase. A typical configuration in the AA phase is shown in Fig.~\ref{cpsl6}.
Our observation of a bimodal probability distribution for the order parameter $\phi$
(see Fig.~\ref{dist}) confirms
the existence of the AA phase.
Although MFT predicts a symmetry-breaking transition in any number of dimensions,
we do not expect such a phase transition in
one dimension; simulations on a ring confirm this.

\begin{figure}[h]
\includegraphics[clip,angle=0,width=0.6\hsize]{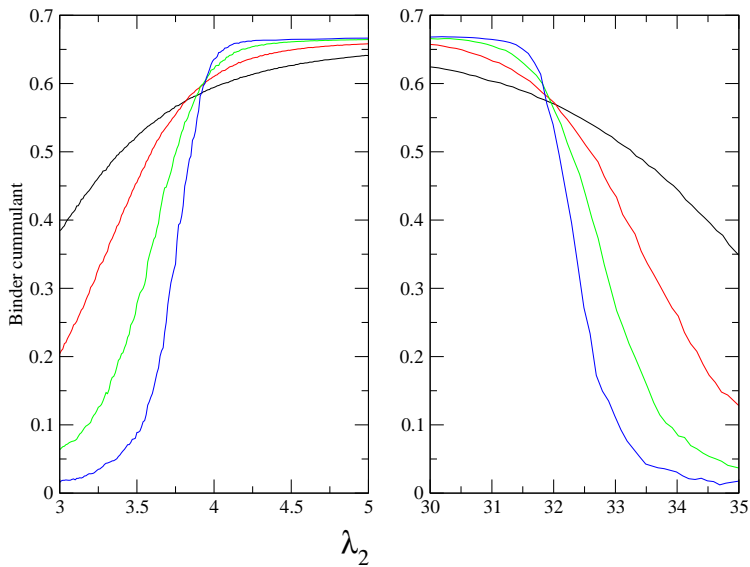}
\caption{{\footnotesize (Color online) Binder cumulant versus $\lambda_2$,
for $\mu=2.0$ and $\lambda_1=0.2$. System sizes: $L=20,40,80,160$ (in order of increasing steepness).}}
\label{binder}
\end{figure}

In order to classify the AA-AS transition, we perform a finite-size scaling study.
First we investigate the behavior of the reduced Binder cumulant \cite{bind81}, given by
\begin{equation}
U_4=1-\frac{\langle{\phi^4}\rangle}{\langle{\phi^2}\rangle^2}.
\end{equation}
The intersection points of the cumulants for successive pairs of sizes depend rather weakly on the sizes,
providing a good estimate for the critical point.
The value of the cumulant at the critical point approaches a universal value as $L\to \infty$.
In Fig.~\ref{binder} we show the results for the case
$\mu=2.0$ and $\lambda_1 = 0.2$. The curves for different sizes
intersect at $\lambda_2 = 3.940(5)$, and again at $\lambda_2 = 31.92(6)$ when $L\to \infty$.

Extrapolating to infinite size, we estimate
$U_{4,c}=0.605(10)$ at the transitions, consistent with the $U_{4,c} =0.61069...$ \cite{kami93}
for the two-dimensional Ising model with fully periodic boundary conditions.
For values of $\lambda_2$ between the transitions points, the cumulant approaches
2/3, as expected in an ordered phase that breaks a ${\cal Z}_2$ (i.e., up-down) symmetry.

\begin{figure}[h]
\includegraphics[clip,angle=0,width=0.5\hsize]{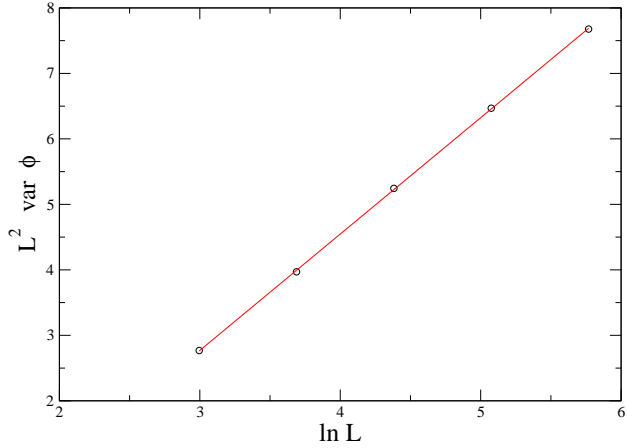}
\caption{{\footnotesize (Color online) Scaled order parameter variance
ln$L^2\,$var$\phi$ versus ln$L$ at the AS-AA transition.
System sizes: $L=20,40,...,320$. Parameters: $\mu=2.0$, $\lambda_1=0.2$, $\lambda_2=3.94$.
The slope of the regression line is 1.76.}}
\label{var}
\end{figure}

Further evidence for Ising-like behavior is furnished by the order-parameter variance.
At the critical point, var$(\phi) =\langle \phi^2 \rangle - \langle\phi\rangle^2$, is expected to scale
so: $L^2$var$(\phi) \propto L^{\gamma/\nu}$; this is confirmed in Fig.~\ref{var}. A fit to the data yields
the exponent ratio ${\gamma/\nu}=1.76(5)$, again in good agreement with the known value of
${\gamma/\nu}=7/4$ for the Ising model in two dimensions.

\begin{figure}[h]
\includegraphics[clip,angle=0,width=0.5\hsize]{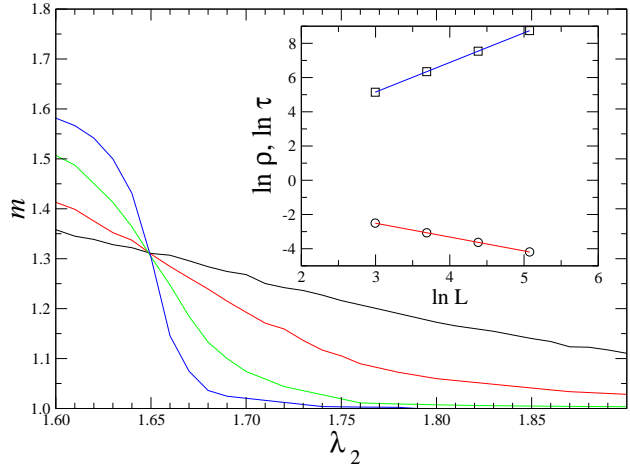}
\caption{{\footnotesize (Color online) Quasistationary moment ratio $m$ versus $\lambda_2$, for
$\mu=2.0$ and $\lambda_1=0$.
System sizes: $L=20,40,80,160$ (in order of increasing steepness). Inset: Scaled critical QS density of active sites ln $\rho$ (bottom)
and scaled lifetime of the QS state ln $\tau$ (top), {\it
versus} ln$L$. Parameters: $\mu=2.0$, $\lambda_1=0$ and $\lambda_2=1.6488$}}
\label{lambdaz-dp}
\end{figure}

\begin{figure}[h]
\includegraphics[clip,angle=0,width=0.5\hsize]{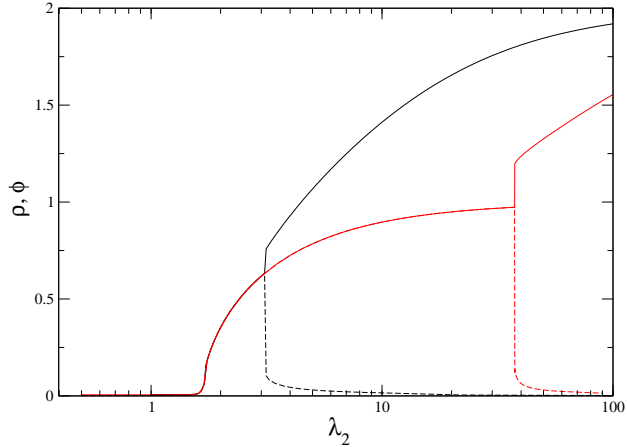}
\caption{{\footnotesize (Color online) Order parameters $\rho$ (solid) and
$\phi$ (dashed) {\it versus} $\lambda_2$,
for $\lambda_1 = 0$. Parameters: $\mu=0.2$ and $\mu=2.0$ from top to bottom. System size: L=80.}}
\label{lambdaz}
\end{figure}

For the $\lambda_1=0$, simulations show no evidence of the asymmetric-active-II phase predicted by
MFT.  Simulations reveal only two phase transitions along this line. The first, from the absorbing to the
asymmetric active phase, occurs, as expected, at the critical value of
the basic CP on the square lattice. We find $\lambda_{2,c}=1.6488(3)$,
and confirm that this transition belongs to the DP class (see Fig.~\ref{lambdaz-dp},
where we find $m=1.320(8)$, $\beta/\nu_\perp=0.81(1)$ and $\nu_{||}/\nu_\perp=1.74(3)$,
consistent with the values for DP in 2+1 dimensions, cited above).
The second transition, between the active-asymmetric and the active symmetric phases,
is {\it discontinuous}. Figure~\ref{lambdaz} illustrates such transitions for $\mu=0.2$ and $\mu=2$.
These results signal a qualitative failure of mean-field theory along the line $\lambda_1=0$.

\section{Conclusions}

A contact process with creation at both nearest and next-nearest
neighbors, and suppression (in the form of an increased annihilation
rate) at neighbors of active sites is shown to exhibit a broken-symmetry
phase with sublattice ordering.
The existence of a continuous symmetry-breaking transition is
predicted by mean-field theory and confirmed in simulations on the
square lattice.
The asymmetric active phase exists for a restricted range
of the second-neighbor creation rate, $\lambda_2$, when that
at first neighbors, $\lambda_1$, is sufficiently small.  For a given
value of the suppression parameter $\mu$, there is a certain value, $\lambda_1^*$,
above which no sublattice order occurs.
We note that the possibility of breaking $Z_2$ symmetry depends on the
fact that creation rates $\lambda_1$ and $\lambda_2$ apply to different sublattices.
We would not expect to observe this symmetry breaking if, for example, there were
a creation rate $\lambda_1$ for {\it both} nearest- and second-neighbors, and a
different creation rate at {\it third-neighbor} sites.

For $0 < \lambda_1 < \lambda_1^*$, the sequence of phases observed on
increasing $\lambda_2$ from small values,
while holding $\mu$ and $\lambda_1$ fixed, is: absorbing, symmetric-active, asymmetric-active,
and symmetric-active.  The first transition belongs to the
universality class of directed percolation, while the latter two are Ising-like.
For the special case of $\lambda_1=0$, the first transition is from absorbing
to asymmetric-active and the second is discontinuous. For both values of $\mu$ studied
we observe continuous transitions for $\lambda_1>0$.
While increasing $\lambda_1$ makes the transition smoother,
we can not exclude the possibility of a discontinuous transition when $\lambda_1 >0$, for
larger values of $\mu$. A complete characterization of the reentrant transition
for $\mu>2.0$, including the location of the possible tricritical points
separating the line of second order transitions from that of first order transitions
is left for future work.

The absorbing-active transition at $\lambda_1=0$ marks the
meeting of two lines of critical points, one of DP transitions, the other of Ising
transitions.  The meeting of two such lines has been studied in the context of a
nonequilibrium Potts model by Droz at al. \cite{droz}, and
of the generalized voter model (GVM) by Al Hammal et al. \cite{alhammal}.
The former study also finds an absorbing state in the DP class, while in the
latter, after the Ising and DP lines join, the absorbing transition belongs
to the GVM class.  We note that in these studies the models have but two (symmetric)
absorbing states (i.e., all plus and all minus), whereas for $\lambda_1=0$, our model has two
(symmetric) absorbing {\it subspaces} (activity restricted to one sublattice) and a third,
completely inactive absorbing state.
In terms of long-time behavior, the absorbing-active phase transition takes
the system from the fully inactive state to one with activity restricted to
one sublattice.
Staring from all sites active, at the critical point
$\lambda_1 = 0$, $\lambda_2 = \lambda_{c,CP}$,
fluctuations drive local regions into states with activity in only one
sublattice, as well as fully inactive regions. We expect that these domains then coarsen, so
that eventually there is activity only in one sublattice; from then on the dynamics is that
of the basic contact process.  Thus it is not surprising that we observe DP-like
{\it static} behavior.  It is nevertheless possible that the short-time critical behavior
includes a regime dominated by domain dynamics as in the GVM.
It also seems possible that the discontinuous AA-AS transition observed for $\lambda_1 = 0$
belongs to the voter model class.
We intend to explore these questions in future work.

Possible extensions of this work include precise determination of static and dynamic critical properties
at the symmetry-breaking transition, studies of the effect of diffusion and/or anisotropy (which may
generate modulated phases) and studies of a model in continuous space.
The latter appears to be of particular interest in the context of
patterns arising in ecosystems or bacterial colonies.
We note that the suppression of activity at the nearest neighbors of active sites
resembles biological lateral inhibition, known to be important in the visual system
of many animals \cite{lytton}.  Thus extensions of the present study might be useful in the
study of patterned activity on the retina.
Finally, if the inhibition rate were a decreasing function
(linear or quadratic) of the number of occupied nearest neighbors (i.e., $\mu < 0$),
one should observe symbiotic coexistence of activity in the two sublattices.
The latter is reminiscent of the
Janzen-Connel hypothesis for the maintenance of tree species biodiversity
in tropical rainforests \cite{janzen-connell}.
\vspace{1cm}

\noindent{\bf Acknowledgments}

This work was supported by CNPq and FAPEMIG, Brazil.  We thank Hugues Chat\'e for
helpful comments.

\bibliographystyle{apsrev}

\newpage

%\begin{figure}[h]
%\epsfysize=10cm \epsfxsize=12cm \centerline{ \epsfbox{CPSLM2.EPS}}
%\caption{\footnotesize{Phase diagram in $\lambda_1 - \lambda_2$ plane
%for $\mu = 2$, showing absorbing (ABS), active-symmetric (AS) and active asymmetric (AA) phases.}}
%\label{cpslm2}
%\end{figure}

%\begin{figure}[h]
%\label{config}
%\epsfysize=10cm \epsfxsize=12cm \centerline{ \epsfbox{cpsl6.eps}}
%\caption{{\footnotesize A typical configuration observed in the active asymmetric phase.
%Particles are represented by black sites. System size: $L=50$. Parameters: $\mu=2.0$,
%$\lambda_1=0.2$ and $\lambda_2=6.0$.}}
%\end{figure}

\end{document}